\documentclass{template/bmcart}

\usepackage{amsthm,amsmath}
\usepackage{amssymb}
\RequirePackage{hyperref}
\usepackage[utf8]{inputenc} 

\usepackage{graphicx}
\usepackage{subcaption}
\usepackage{tabularx}
\usepackage[algo2e,ruled,lined,boxed]{algorithm2e}

\startlocaldefs
\endlocaldefs

\begin{document}

\begin{frontmatter}

\begin{fmbox}
\dochead{Research}


\title{Beyond the Surface: Advanced Wash Trading Detection in Decentralized NFT Markets}


\address[id=aff1]{
  \orgdiv{The Faculty of Mathematics, Natural Sciences and Information Technologies},             
  \orgname{University of Primorska},          
  \street{Glagoljaška 8},
  \city{Koper},                              
  \cny{Slovenia}                                    
}
\address[id=aff2]{%
  \orgname{InnoRenew CoE},
\street{Livad 6a},
  \city{Izola},
  \cny{Slovenia}
}

\author[
  addressref={aff1,aff2},
  email={aleksandar.tosic@upr.si}
]{\inits{A.T.}\fnm{Aleksandar} \snm{Tošić}}

\author[
  addressref={aff1,aff2},
  email={niki.hrovatin@famnit.upr.si}
]{\inits{N.H.}\fnm{Niki} \snm{Hrovatin}}

\author[
  addressref={aff1},
  email={jernej.vicic@upr.si}
]{\inits{J.V.}\fnm{Jernej} \snm{Vičič}}





\end{fmbox}


\begin{abstractbox}

\begin{abstract} 
Wash trading in decentralized markets remains a significant concern magnified by the pseudonymous and public nature of blockchains. In this paper we introduce an innovative methodology designed to detect wash trading activities beyond surface-level transactions.
Our approach integrates NFT ownership traces with the Ethereum Transaction Network, encompassing the complete historical record of all Ethereum account normal transactions. 
By analyzing both networks, our method offers a notable advancement over techniques proposed by existing research.
We analyzed the wash trading activity of 7 notable NFT collections.
Our results show that wash trading in unregulated NFT markets is an underestimated concern and is much more widespread both in terms of frequency as well as volume. 
Excluding the Meebits collection, which emerged as an outlier, we found that wash trading constituted up to 25\% of the total trading volume. Specifically, for the Meebits collection, a staggering 93\% of its total trade volume was attributed to wash trading.

\end{abstract}


\begin{keyword}
\kwd{NFT}
\kwd{Washtrading}
\kwd{Network Analysis}
\kwd{Blockchain}
\end{keyword}


\end{abstractbox}
%

\end{frontmatter}




\section{Introduction}
Non-fungible tokens, commonly referred to as NFTs, have ushered in a fresh paradigm within the realm of financial assets. At their core, NFTs signify digital manifestations of ownership rights, securely recorded through unalterable contracts linked to unique addresses or hashes on a blockchain. Although this underlying principle holds broad applicability, spanning domains like supply chains, real estate, and music, NFTs have predominantly captivated the financial landscape when employed to confer digital ownership over visual content (images) or videos.

To facilitate trading of NFTs, decentralized markets were created using a set of smart contracts. These contracts govern the exchange of ownership between two parties (wallets) for a certain price. Likewise, the transfer of ownership can occur without a trade as well. The fast growth of these marketplaces has led many to question the legitimacy of market activities given the questionable fundamentals of the underlying assets the NFTs represent. These concerns are strengthened with examples of single NFTs being traded for an unrealized profit of over 3 million USD. \cite{SERNEELS2023103374}, which is a result of wash trading.

Wash trading refers to the deceptive practice of artificially inflating trading activity by conducting trades between parties that are controlled or colluded by the same entity. In this scheme, the buyer and seller appear to be separate entities engaging in legitimate transactions, but in reality, they are orchestrated by the same individual or group. Wash trading can lead to misleading market signals, false perceptions of demand, distorted price levels, and the creation of an illusion of liquidity. This unethical activity entices other market participants to trade based on inaccurate information, ultimately compromising the integrity and efficiency of the market.
Wash trading has been widely recognized as a detrimental practice in conventional financial markets due to its potential to deceive and manipulate. For instance, in the United States, regulations such as the Commodity Exchange Act (CEA) of 1936 explicitly prohibit wash trading to maintain the fairness and transparency of financial markets.

In the context of the cryptocurrency ecosystem, wash trading remains a concern, especially in unregulated sectors such as the crypto markets and more specifically, non-fungible token (NFT) trading networks. NFTs, being digital assets represented on blockchains, have seen instances of wash trading that aim to artificially boost the perceived value of these assets. This practice is particularly challenging to address because blockchain transactions are publicly recorded and accessible, but wallet addresses involved in transactions lack personal identifying information, making it difficult to ascertain the true parties behind the trades.
The emergence of wash trading in NFT markets highlights the need for effective detection and prevention mechanisms to maintain the credibility and legitimacy of trading activities. Proposed algorithms and techniques aimed at identifying wash trading in pseudonymous networks like blockchain-based NFT trading can significantly contribute to mitigating the negative impacts of this practice and ensuring a more transparent and trustworthy trading environment.

According to a study published by the National Bureau of Economic Research \cite{cong2022crypto}, wash trading attributed 70\% of the volume in cryptocurrency trading. Additionally, Dune, a company focusing on cryptocurrency market analysis, published a report claiming up to 80\% of volume in NFT markets was wash traded peaking in January of 2022 \cite{dune}. The surprising growth of the market coupled with questionable fundamentals highlights the need for a reliable method for detecting wash trading for market participants, entrepreneurs, stakeholders and most importantly regulators.

\section{Literature Review}
\label{sec:lit-review}
NFT markets are a fairly new phenomenon and while the academic community is not agnostic to the enormous growth in popularity, methods for detecting illicit activity have been lacking. Additionally, existing methods from traditional finance are not portable to blockchain based assets due to the pseudonymous nature of market participants. Detecting wash trading in NFT markets can be categorized in two distinct approaches. Identification and quantification through data analysis, which largely depends on statistical models for identification, and graph theory approaches, where the goal is to search large trading graphs for patterns that may be a result of wash trading. The NFT market is primarily driven by the cryptocurrency market, it is reasonable to assume that some fraudulent activities already discovered in cryptocurrencies are transferable to the NFT market~\cite{ante2022non}.

Cao et al.~\cite{cao2016} were among the first to analyse wash trading in traditional markets by defining trading patterns and directed graphs on order book data to identify wash trading.
Tariq and Sifat~\cite{tariq2022suspicious} first presented a work on statistical analyses (mainly Benford’s law and clustering effects) to show abnormal prices in automated trades. The work is further extended and filtered by almost the same authors in  \cite{sifat2023suspicious}. They observe general nonconformity to Benford's Law in a manner possibly indicative of price manipulation.

The following approaches for discovery of potential wash trading used methodologies based on construction of directed graphs of transactions and searching for cycles or groups of addresses that were potentially involved in wash trades. 
von Wachter et al.~\cite{vonwachter2022nft} observe 3.5 year span NFT trading looking for wash trades. They report that almost 4 \% of all addresses triggered potentially illicit trading patterns. They constructed directed graphs in the NFT network and observe topological cycles expressing wash trades.
Liu et al.~\cite{liu2023nft} uses a similar approach augmented with different heuristic based on time component to mellow down the complexity of the problem. They report reveals that on average, 0.14 \% of transactions, 0.11 \% of wallets, and 0.16 \% of tokens in each collection were involved in wash trading on a test sample of 7 popular NFT collections. The final limitation that they present in the paper, the unknown wallet from the ETH network, has been addressed in our approach.
Sven Serneels~\cite{serneels2023} also presents a similar methodology focusing on token networks, proposing three types of candidates for wash trading: closed loop token trades, closed loop value trades, and high transaction volumes.

All presented methods based on directed graphs have an important drawback by focusing only on transactions that directly involve buying/selling or moving NFT tokens. These methods are not able to discover wash trades, where part of the transactions occur on the Ethereum network. The size of the whole Ethereum network, even considering a narrower time-frame, makes all presented approaches unfeasible




\section{Wash Trading Definition and the Challenges in Decentralized Markets}
\label{sec:wash-definition}
The wash trading practice is rooted in the traditional finance, it was first prohibited in the U.S. with the 1936 Commodity Exchange Act~\cite{us_wash-trade}. A well known definition is given by the Commodity Futures Trading Commission (CFTC) "Entering into, or purporting to enter into, transactions to give the appearance that purchases and sales have been made, without incurring market risk or changing the trader's market position."~\footnote{CFCT wash sale definition: \url{https://www.cftc.gov/LearnAndProtect/AdvisoriesAndArticles/CFTCGlossary/index.htm\#washtrading}}

A wash trade is a form of market manipulation in which an investor simultaneously sells and buys the same financial instruments, creating misleading, artificial activity in the marketplace. This can give a false impression of volume or market interest.

The advent of decentralized markets and platforms has added complexity to the landscape of wash trading. Decentralized finance (DeFi) platforms operate without centralized intermediaries, which means traditional enforcement mechanisms might not be directly applicable. 
Furthermore, the inherent decentralization and openess of blockchains permit users to generate an unlimited number of accounts, identifiable only by cryptographic addresses, without any reliable means to verify if multiple accounts are under single ownership. 
While the transparency of blockchains ensures every transaction is visible, it paradoxically aids wash trading, as the false volume is readily observable to the wider market, potentially misleading others about the true demand or interest in an asset.
In addition, the diversity of DeFi platform implementations makes tracking wash trading across these ecosystems increasingly complex.

Nonetheless, the defining characteristics of wash trading remain consistent, even in the decentralized space: a series of transactions that, upon completion, leave the trader in the same market position as before. The intent, as always, is to give the illusion of genuine market activity without incurring actual risk. In decentralized markets, evidence suggests that wash trading is facilitated by the transparency and pseudonymity inherent in these systems.

\section{Methodology}

As highlighted in Section~\ref{sec:lit-review}, the current literature seeks to uncover wash trading by analyzing the NFT ownership traces, which capture the history of ownership changes an NFT undergoes from its creation to its current state. By observing NFT transfers and analyzing these shifts in ownership, researchers aim to recognize potentially suspicious patterns. Though this approach has flagged some wash trading activities, it likely only uncovers a fraction of the broader issue. 

The intricacies of wash trading are deeply woven into the blockchain, especially due to its public and pseudonymous nature. Such characteristics make it relatively easy for malicious actors to elude current wash trading detection solutions.
The repercussions of NFT wash trading can be profound. It has the potential not just to skew the value of individual NFTs but also to inflate the perceived worth of entire NFT collections, and by extension, sway the broader cryptocurrency market dynamics.

In contrast to existing approaches, our research aims to detect wash trading patterns by integrating NFT ownership traces with a derived structure from the transaction histories of accounts engaged in NFT trading. This structure, which we refer to as the Linkability Network, captures inherent connections and potential collaborations within the blockchain. Through this integration, we can better discern relationships between accounts, whether they are controlled by a single entity or are in collusion to artificially inflate the value of specific assets, thus improving our capability to detect wash trading.

The term Linkability Network has been purposefully chosen to reflect the objective and methodology of our study.  In the context of blockchains, accounts act as pseudonyms, with a single user capable of holding multiple accounts. Direct p2p transactions of native blockchain currency imply a relationship of trust between the two accounts involved. Therefore, indicating that the two accounts—and by extension, the users managing these accounts—are related in some way. According to Pfitzmann and Hansen's comprehensive terminology work on data minimization and privacy~\cite{pfitzmann2010terminology}, 'linkability' refers to the capacity to sufficiently distinguish whether two or more items of interest are related within the system. In our scenario, the items of interest are the blockchain accounts. Consequently, the Linkability Network effectively illustrate the potential connections or 'links' between accounts as inferred from the whole transaction history.

To construct the Linkability Networks, we initially procured NFT ownership traces, as given in Section~\ref{sec:nft-traces}, and we extracted the complete Ethereum Transaction Network from an Ethereum full node to obtain the transaction history of accounts of interest, as outlined in Section~\ref{sec:eth-network}. Section~\ref{sec:linkability-net} elaborates on the construction of the Linkability Networks. Our methodology for wash trading detection stands on the integration of NFT ownership traces with the Linkability Networks, a process which we detail in Section ~\ref{sec:wash-trade-metod}.

\subsection{Data Collection and Cleaning}

\subsubsection{NFT Ownership Traces}
\label{sec:nft-traces}

To narrow down our analysis to only relevant NFT collections, we employed the opensea-scraper~\footnote{OpenSea Scraper: \url{https://github.com/dcts/opensea-scraper}} npm package to fetch metadata of top NFT collections from the OpenSea platform. This metadata guided our selection process, allowing us to focus on significant collections when validating our methodology.
During our selection process we decided to consider only NFT on the Ethereum network as it is the most popular NFT enabled blockchain. We prioritized top collections based on monetary value, recognizing that these collections see the majority of fund transfers. For manageability, we narrowed our scope to collections containing fewer than 20,000 NFTs. Furthermore, we consider only the PFP (Profile Picture tokens) category, reflecting its soaring popularity within the blockchain community. Lastly, our research was restricted to ERC721 tokens. According to the above criteria, the following collections were selected to validate our study: bored ape yacht club, mutant ape yacht club, azuki, cloonex, meebits, bored ape kennel club, beanz official.

NFT ownership traces were obtained by querying the Etherscan APIs. 
We used the logs module of the Etherscan APIs to obtain transfers of NFT ownership by specifying the collection contract address, the method signature of the NFT ownership transfer, and querying the endpoint from the NFT contract creation date, 1000 Ethereum blocks at a time until 1/5/2022.

The obtained NFT ownership traces were enriched by integrating data on the value transfers corresponding to any closed auctions tied to the NFT transfers. This auxiliary data was generously supplied by OpenCloset.ai and encompasses auctions finalized on the following NFT marketplaces: Blur, OpenSea, Looks Rare, X2Y2. Value transfers from other NFT marketplaces were incorporated using the OpenSea API. As a result, the data we obtained should encompass all NFT ownership changes and nearly all value transfers of NFT trades.

We define an \textbf{NFT ownership trace} as a directed weighted temporally ordered transaction graph, denoted as $G_{nft} = (V, E, w,T)$ where:

\begin{itemize}
    \item $V\subseteq V$ is the set of vertices, each representing an EOA that has owned a specific NFT from a collection.
    \item $E$ is the set of directed edges, with each edge $(v, u) \in E$ representing the NFT's transition from the transferor $v$ to the transferee  $u$. 
    \item $w: E \rightarrow \mathbb{R_{\ge 0}}$ is a weight function that assigns to each edge $(v, u)$ a weight equal to the value transfer ( from $u$ to $v$) at an NFT trade or a weight of 0 to signal an simple NFT ownership transfer.
    \item $T:E \rightarrow \mathbb{R}$ is a time function, mapping each edge to a real number that represents the timestamp of the transfer. For any two edges \( e_1, e_2 \in E \) such that \( e_1 \) precedes \( e_2 \), it holds that \( T(e_1) < T(e_2) \), ensuring temporal order.
\end{itemize}

\subsubsection{Ethereum Transaction Network}
\label{sec:eth-network}

As a foundation for constructing linkability networks, we needed to obtain the transaction history of accounts involved in NFT trading. Specifically, we are interested only in normal transactions. In the Ethereum context, a "normal transaction" refers to a simple transfer of Ether (ETH) from one externally owned account (EOA) to another, without the invocation of a smart contract's function. Such transactions suggest that the involved accounts may either belong to the same entity or share a significant mutual trust relationship.

To model this transactional relationship, we represent the \textbf{Ethereum Transaction Network} as a directed graph, denoted as $G_{eth} = (V, E)$, where:

\begin{itemize}
    \item $V$ is the set of all unique Ethereum EOA accounts, each account being a vertex in the graph.
    \item $E$ is the set of all directed edges, where each edge $(v, u) \in E'$ represents a normal transaction of ETH currency from account $v$ to account $u$.
\end{itemize}

In this graph, the direction of an edge signifies the direction of the transaction, from the sender to the receiver. The Ethereum Transaction Network denotes the history of normal transactions of all externally owned accounts spanning the entirety of Ethereum's existence up to its transition to proof-of-stake consensus on 15.09.2022 at block: \#15537393.

To obtain these data we synched a go-ethereum~\footnote{Go-Ethereum: \url{https://geth.ethereum.org/}} full node to the Ethereum mainnet, and we ran the ethereum-etl~\footnote{Ethereum-etl: \url{https://ethereum-etl.readthedocs.io/en/latest/}} tool for extracting all Ethereum transactions. The extraction process took nine days, producing 980GB of output data on a dedicated server~\footnote{100 cores and 1TB of memory. Most of the memory was used.}.
The large output file was then filtered using grep to keep only transactions having input 0x and value $>$ 0. Therefore, we filter out all smart contract transactions and value transfers with 0 amount. The filtering took 22 hours resulting in a file of 188GB.
The resulting file was further filtered to remove all transactions involving addresses of the top centralized exchanges and well-known mining pools. The list of removed addresses spans more than 800 entries and is available at~\footnote{Dune analytics spellbook: \url{https://github.com/duneanalytics/spellbook/tree/main/models/cex}}.

\subsection{Linkability Network}
\label{sec:linkability-net}

To make the extensive information of the Ethereum Transaction Network more manageable, we devised the concept of the Linkability Network.

The \textbf{Linkability Network} is a directed, weighted graph, denoted as $G_{ln} = (V', E')$, derived from the Ethereum Transaction Network $G_{eth} = (V, E)$, where:

\begin{itemize}
    \item $V'\subseteq V$ is the set of vertices, each representing a unique account that has owned an NFT from a specific collection at any point in time.
    \item $E'$ is the set of directed edges, with each edge $(v, u) \in E'$ representing the shortest path of ETH currency transfer from $v$ to $u$ in the Ethereum Transaction Network. 
    \item $w: E' \rightarrow {1, 2, ..., d}$ is a weight function that assigns to each edge $(v, u)$ a weight equal to the length of the shortest path from $v$ to $u$ in the Ethereum Transaction Network.
\end{itemize}

Therefore, an edge $(v,u)$ with $w_{v,u}=2$ in a linkability network shows that in the Ethereum Transaction Network there is a path of normal transactions of length 3 between node $v$ and $u$, e.g. ($v \rightarrow a \rightarrow b \rightarrow u$). 
It is important to note, that these normal transactions imply a relationship of trust or a shared agreement between the two accounts involved. In the absence of contractual stipulations or automated logic, as with smart contracts, both parties are adhering to some shared agreement or understanding outside the purview of automated contract enforcement. This adherence to their mutual arrangement suggests a deeper trust relationship or shared control. As a result, there is a much higher probability that the two accounts participating in a normal transaction are either owned or controlled by the same entity, or entities with a strong mutual trust relationship. 

The edge weight in a linkability network is thus inversely related to the strength of the collaboration or trust connection between the edge's vertices. Therefore, when constructing linkability networks, it's pragmatic to set a reasonable limit on the search depth for the shortest path. This not only expedites the construction process but also reduces memory usage and prevents the saturation of the linkability network.

\subsubsection{Linkability Network Construction}
\label{sec:linkability-net-construction}
Linkability networks were constructed with a Java program using the JGraphT~\footnote{JGraphT: \url{https://jgrapht.org/}} library. 
While the JGraphT library offers robust functionality, it doesn't support concurrent execution for its all-shortest-path algorithms. 
Given the vast volume of data we're handling, a single-threaded approach would prove highly time-consuming. To expedite the processing and make optimal use of our multicore server, it was imperative to employ concurrent operations. Moreover, our specific requirements necessitated limiting the source and target vertex set to a particular list of addresses, a customization that the default algorithms couldn't readily accommodate. As a result, we opted to implement the construction of linkability networks by running multiple instances of the BFS (Breadth-First Search) algorithm concurrently.

A linkability network was constructed for each of the selected NFT collections by running the algorithm given in~\ref{alg:bfs}.
Initially, the Ethereum transaction network was read in memory as a directed graph with respect to the ETH token transfer direction, and addresses owning NFTs of a selected collection were loaded. 
Then, each NFT owner address was used as a root vertex for a Breadth First Search (BFS) operating on the Ethereum transaction network. 
The BFS described in~\ref{alg:bfs} perform a traversal up to the specified maximal depth, adding edges to the output graph from the root vertex to a visited vertex if the visited vertex is in the set of addresses owning NFTs of the selected collection. The visit depth is integrated in the output graph as the edge weight.

For each selected NFT collection, a linkability network was generated using the algorithm detailed in~\ref{alg:bfs}. 
The process began by loading the Ethereum transaction network $G_{eth}$ into memory as a directed graph with respect to the ETH token transfer direction, and $A$ the set of addresses owning NFTs of the current collection under consideration were loaded.  
Subsequently, each NFT owner address $a \in A$ served as a root vertex for a Breadth First Search (BFS) on the Ethereum transaction network. As described in~\ref{alg:bfs}, this BFS traverses up to a predefined maximum depth, incorporating edges into the output graph when the traversed vertex $v \in A$. The depth of the visit determines the edge weight in the output graph.

\begin{algorithm2e}
\footnotesize
\KwIn{Ethereum transaction network $G_{eth}$, Set of NFT collection owner addresses $A$, Max depth $D$}
\KwOut{Linkability Network $G_{ln}$}
\ForEach{address $a$ in $A$ \textup{\textbf{in parallel}}}{
    Initialize empty set $Visited$ and add $a$ to $Visited$Text for this section\ldots;\\
    Initialize queue $Q$ and enqueue $a$ and enqueue $null$;\\
    Initialize empty graph $G_a$;\\
    $depth = 0$;\\
    \While{Q is not empty}{
        $v$ = dequeue $Q$;\\
        \If{$v == null$}{
            $depth++$;\\
            $Q$ enqueue $null$;\\
            \If{$peek(Q) == null$}{break;}
            continue;
        }
        \If{$depth < D$}{
            \ForEach{neighbor $u$ of $v$ in $G$}{
                \If{$u$ not in $Visited$}{
                    Enqueue $u$ in $Q$;\\
                    Add $v$ to $Visited$;\\
                    \If{address $v$ in $A$}{
                        Add edge $e(a,v,depth)$ to $G_a$;
                    }
                }
            }
        }
    }
    \textup{\textbf{Synchronize:}} Merge $G_a$ to $G_{ln}$;
}
\caption{Parallel BFS for Linkability Network Generation}
\label{alg:bfs}
\end{algorithm2e}

\subsection{Wash Trading Detection}
\label{sec:wash-trade-metod}

To detect wash trading, we leverage NFT ownership traces combined with the intricate structure of linkability networks. This approach captures the condensed transaction histories of EOAs involved in NFT trading, enabling us to discern suspicious activities beyond mere surface-level transactions.
To explain the key intuition behind our detection algorithm we recall the wash trading definition given in section~\ref{sec:wash-definition}: executing trades without changing the trader's market position to artificially boost an asset's volume. 
Consequently, such trades occur between interconnected entities. 
To detect potential wash sales, our algorithm traverses the NFT ownership trace. For each NFT trade, it examines if an association exists between the seller and the buyer.
Throughout this paper, we'll refer to this association as a "link", and our algorithm seeks this link within both the NFT ownership trace and the linkability network.

The link within the NFT ownership trace is inferred from changes in NFT ownership. As outlined in section~\ref{sec:nft-traces}, the edge weights in the NFT ownership trace imply two types of NFT ownership transition.
\begin{itemize}
    \item An \textbf{NFT trade}: Indicated by an edge weight greater than 0, implying a value transfer in exchange for the NFT. As these trades are performed on a Decentralized Exchange (DEX), a direct association between the involved parties is typically absent.
    \item An \textbf{NFT transfer}: Represented by an edge weight of 0, this means the NFT was transferred without monetary exchange executed by a smart contract. Such transfers are likely between EOAs controlled by the same entity or as part of off-chain agreements. Given the decentralized and pseudonymous nature of blockchains, establishing external agreements can be risky and challenging. Hence, in both scenarios, an NFT transfer implies a link between the parties involved.
\end{itemize}

Conversely, a link derived from the linkability network is identified by examining the network to check if an edge connects two EOAs. As discussed in section~\ref{sec:linkability-net}, the presence of an edge in the linkability network implies a relationship between the two EOAs EOAs at its endpoints.

Our wash trade detection algorithm utilizes the concept of "link" to cluster EOAs that are potentially involved in deceptive trading practices. The specifics of our approach are detailed in Algorithm~\ref{alg:wash-detection}. Here's a step-by-step breakdown:

\begin{enumerate}
    \item \textbf{NFT Transfer Clustering:} Initially, EOAs linked by NFT transfers (where the edge weight is 0) are grouped together, as seen in the {\footnotesize \textsc{ClusterOnNftTransfer}} function. This results in a list of sets, $L$.
    
    \item \textbf{Merging Sets:} The sets within $L$ with common elements are continuously merged until they are disjoint. This step ensures that EOAs linked by NFT transfers are grouped together.
    
    \item \textbf{Linkability Network Clustering:} To further refine our clusters, we leverage the linkability network. The  {\footnotesize \textsc{ClusterOnLinkabilityNetwork}} function identifies and merges sets in $L$ based on links present in the linkability network.
    
    \item \textbf{Wash Sale Detection:} In the final step, the algorithm iterates over the NFT ownership trace, specifically focusing on NFT trades the ones with the edge weight $>$ 0. A NFT trade is flagged as a potential wash trade if both the seller and buyer addresses are found within the same set in $L$.
\end{enumerate}

This systematic clustering and analysis process enables us to identify suspicious trades, offering insights into possible market manipulation.

\begin{algorithm2e}
\footnotesize

\DontPrintSemicolon

\SetKwFunction{ClusterOnNftTransfer}{ClusterOnNftTransfer}
\SetKwFunction{MergeCommonSets}{MergeCommonSets}
\SetKwFunction{ClusterOnLinkabilityNetwork}{ClusterOnLinkabilityNetwork}
\SetKwFunction{AreAnyLinked}{AreAnyLinked}
\SetKwFunction{Main}{Main}

\KwData{$G_{nft}$ (NFT ownership trace graph), $G_{ln}$ (linkability network)}
\KwResult{List of disjoint sets \(L\), indicating potential wash trade groups}

\BlankLine

\SetKwProg{Fn}{Function}{}{}

\Fn{\ClusterOnNftTransfer{$G_{nft}$}}{
    Initialize an empty list of sets \(L\)\;
    \ForEach{edge \(e\) in \(G_{nft}\)}{
        \If{ weight $w_{e}$ == 0}{
            Add the set \(\{e.source, e.destination\}\) to \(L\)\;
        }
    }
    \Return \(L\)\;
}

\BlankLine

\Fn{\MergeCommonSets{$L$}}{
    \tcp{Merge sets in \(L\) with common elements until all sets are disjoint.}
    \Return \(L\)\;
}

\BlankLine

\Fn{\AreAnyLinked{$S_1$, $S_2$, $G_{ln}$}}{
    \tcp{Check if any element of $S_1$ is linked with any element of $S_2$ in $G_{ln}$.}
    \Return{True/False}\;
}

\BlankLine

\Fn{\ClusterOnLinkabilityNetwork{$L$, $G_{ln}$}}{
    Initialize a flag \(linked = \text{True}\)\;
    \While{linked}{
        Set \(linked = \text{False}\)\;
        \ForEach{ pair of sets \((S_a, S_b)\) in \(L\)}{
            \If{\AreAnyLinked{$S_a$, $S_b$, $G_{ln}$}}{
                Merge \(S_b\) into \(S_a\), and remove \(S_b\) from \(L\)\;
                Set \(linked = \text{True}\)\;
            }
        }
    }
    \Return \(L\)\;
}

\BlankLine

\Fn{\Main{}}{
\(L \leftarrow\) \ClusterOnNftTransfer{$G_{nft}$}\;
\(L \leftarrow\) \MergeCommonSets{$L$}\;
\(L \leftarrow\) \ClusterOnLinkabilityNetwork{$L$, \(G_{ln}\)}\;
\ForEach{edge \(e\) in \(G_{nft}\)}{
    \If{ weight $w_{e}$ \(>\) 0}{and
        \ForEach{set \(S\) in \(L\)}{
            \If{\(\{e.source, e.destination\} \subseteq S\)}{
                \tcp{A wash sale is identified}
            }
        }
    }
}
}
\caption{Detect Wash Trade in NFT Ownership Trace}
\label{alg:wash-detection}
\end{algorithm2e}

\subsection{BFS Depth Selection}
\label{sec:bfs-depth-selection}
In this section, we delve into the rationale behind selecting the maximum depth for BFS traversal during the construction of linkability networks.
The chosen depth directly influences the number of links in the linkability network. Specifically, a deeper traversal tends to enhance the sensitivity of our wash trade detection algorithm, potentially identifying a larger number of suspected wash trades. This increased sensitivity, however, comes with a caveat—it may elevate the false positive rate. Additionally, it's worth noting that deeper traversal also results in extended running times, which can impact the efficiency and scalability of our methodology, especially given that the creation of linkability networks represents the most resource-intensive phase of our approach.

Due to the absence of a controlled dataset of NFT trades with pre-flagged suspicious transactions, we relied solely on evaluating the output of our methodology while judiciously determining an adequate maximum BFS depth.
For this assessment, we designed an experiment to push the BFS depth.
We began by randomly selecting 1,000 NFTs from the 'Bored Ape Yacht Club' collection, subsequently filtering the NFT ownership traces specific to these chosen NFTs. 
Following this, we generated a linkability network for the selected NFTs, setting the maximum depth to 20.
The wash trade detection algorithm~\ref{alg:wash-detection} was executed 20 times for each NFT ownership trace. During each iteration, we incremented a parameter that begins at 1 and maxes out at 20. This parameter directly influences the function {\footnotesize \textsc{AreAnyLinked}}, which we adjusted to seek links that have weights lesser than the given parameter value. 
In this way we can evaluate the output of our approach with a varying BFS maximal depth of up to 20.

\begin{figure}[h]
  \includegraphics[width=0.7\textwidth]{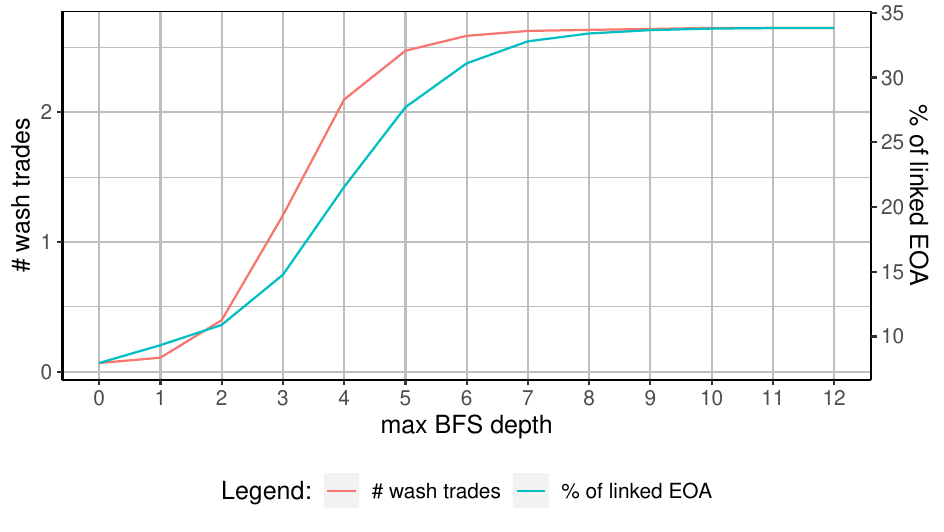}
  \caption{Average number of wash trades and percentage of linked accounts across different BFS max depths. We display only BFS depths up to 12, as beyond this, the response variables remain consistent.}
  \label{fig:dept-test}
\end{figure}

Figure~\ref{fig:dept-test} presents the results of the experiment, illustrating two response variables across various BFS depths: the average number of wash trades and the percentage of linked accounts. The figure indicates that the response variables follow a sigmoidal pattern. Because of this sigmoid shape, our focus shifted to the BFS depth at the inflection point of the average number of wash trades. At this point, the average count of detected wash trades is 1.2, and the percentage of linked accounts stands at 15\%. We find these values to be adequate. 
Consequently, we executed our wash trading detection algorithm on the selected NFT collections, utilizing linkability networks constructed at a BFS depth of 3. This corresponds to a normal transaction path with a length of up to 4.
Obtained results are presented in section~\ref{sec:results}.
Despite the computational intensity associated with traditional network analysis techniques, Table~\ref{tab:computation} underscores the practicality and feasibility of our proposed method.

\begin{table}[]

\begin{tabular}{l|r|r}
\textbf{Operation}             & \multicolumn{1}{c|}{\textbf{Data Output size}} & \multicolumn{1}{c}{\textbf{Time}} \\ \hline
NFT collection scraping        & 260 MB                                         & 13 days                           \\ \hline
Ethereum ETL                   & 988 GB                                         & 9 days                            \\ \hline
Smart contract filtering       & 188 GB                                         & 22 hours                          \\ \hline
Linkability network generation & 10.2 GB                                         & 27 hours                          \\ \hline
\end{tabular}%

\caption{Output size and elapsed time of operations performed to obtain the address linkage graphs.}
\label{tab:computation}
\end{table}

\section{Results}
\label{sec:results}

Unfortunately, validating the results is difficult absent known publicly available cases of wash trading. However, the results clearly highlight the effectiveness of the proposed method and identify cases where the argument of deniability is hard to defend. We present the results objectively and submit them to the reader to judge the results objectively.

\begin{figure}[h!]
\centering
\begin{tabular}{ccc}
\subcaptionbox{\label{azuki}}{\includegraphics[width=0.4\textwidth]{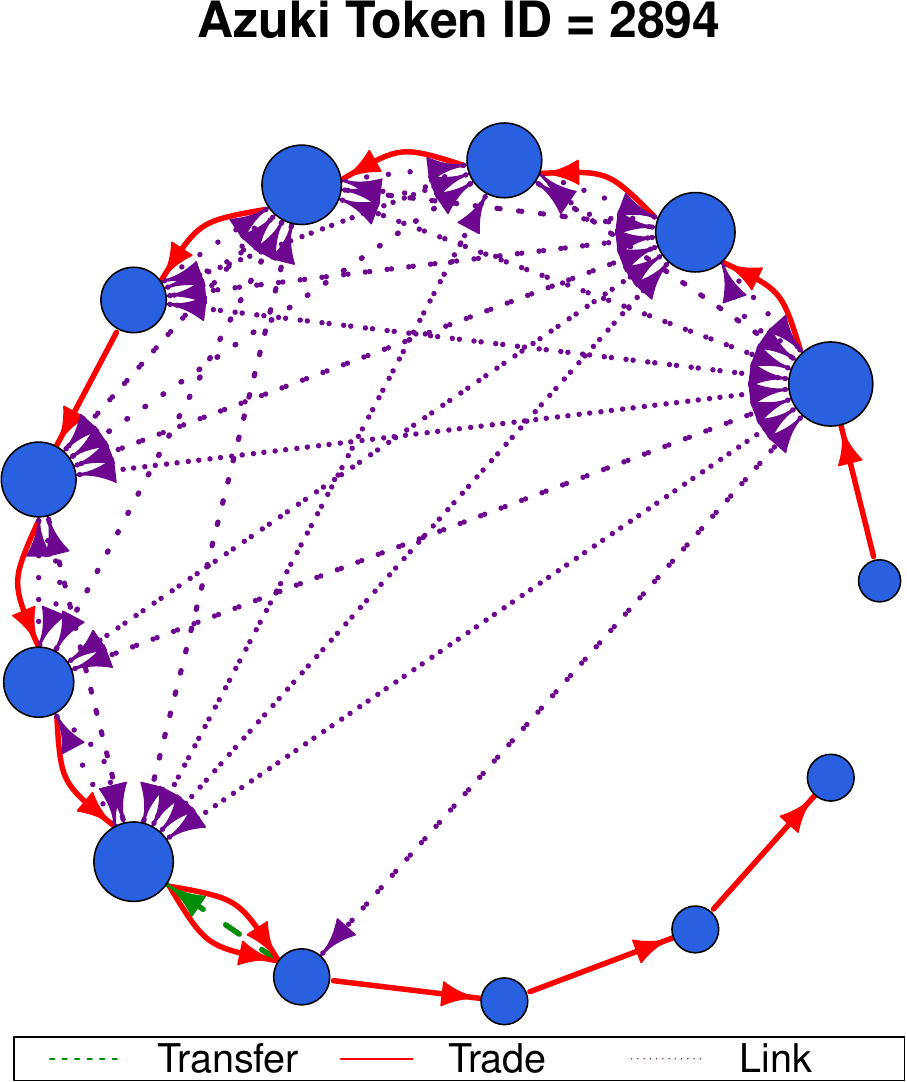}} &
\subcaptionbox{\label{bored}}{\includegraphics[width=0.4\textwidth]{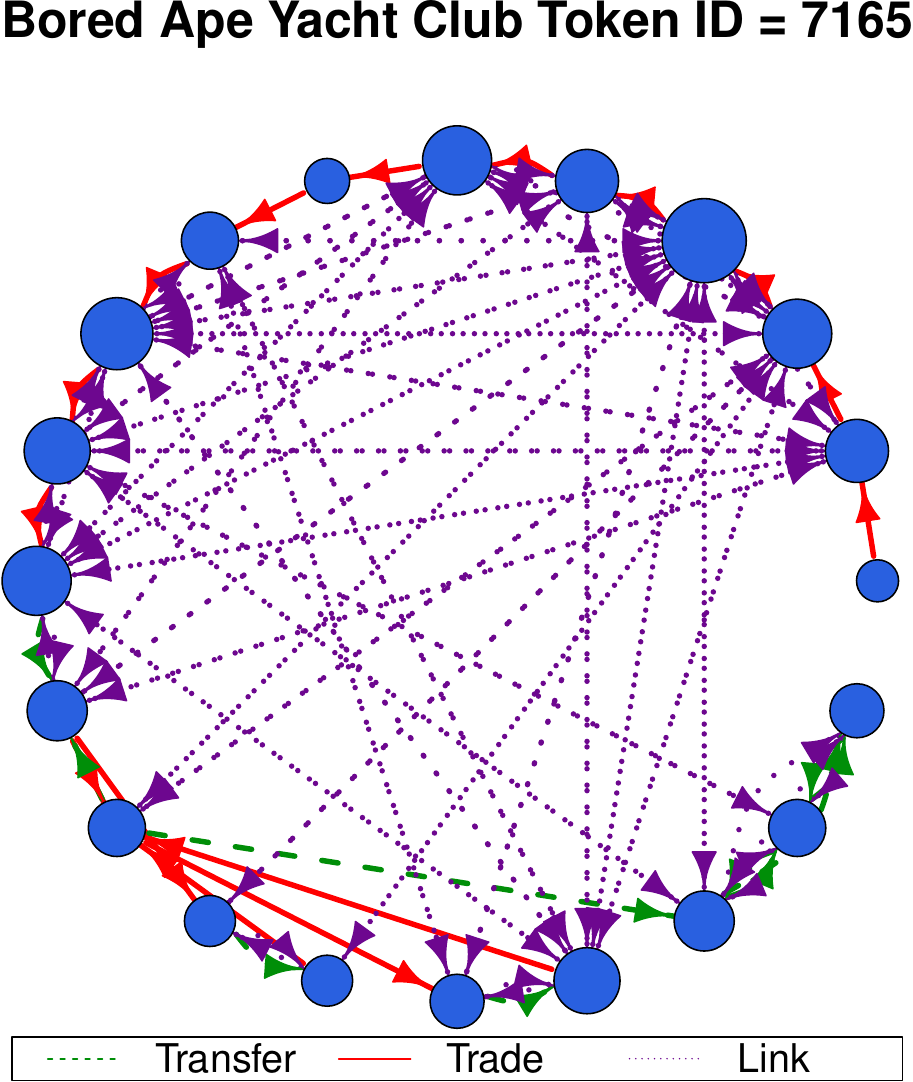}} \\
\subcaptionbox{\label{meebits}}{\includegraphics[width=0.4\textwidth]{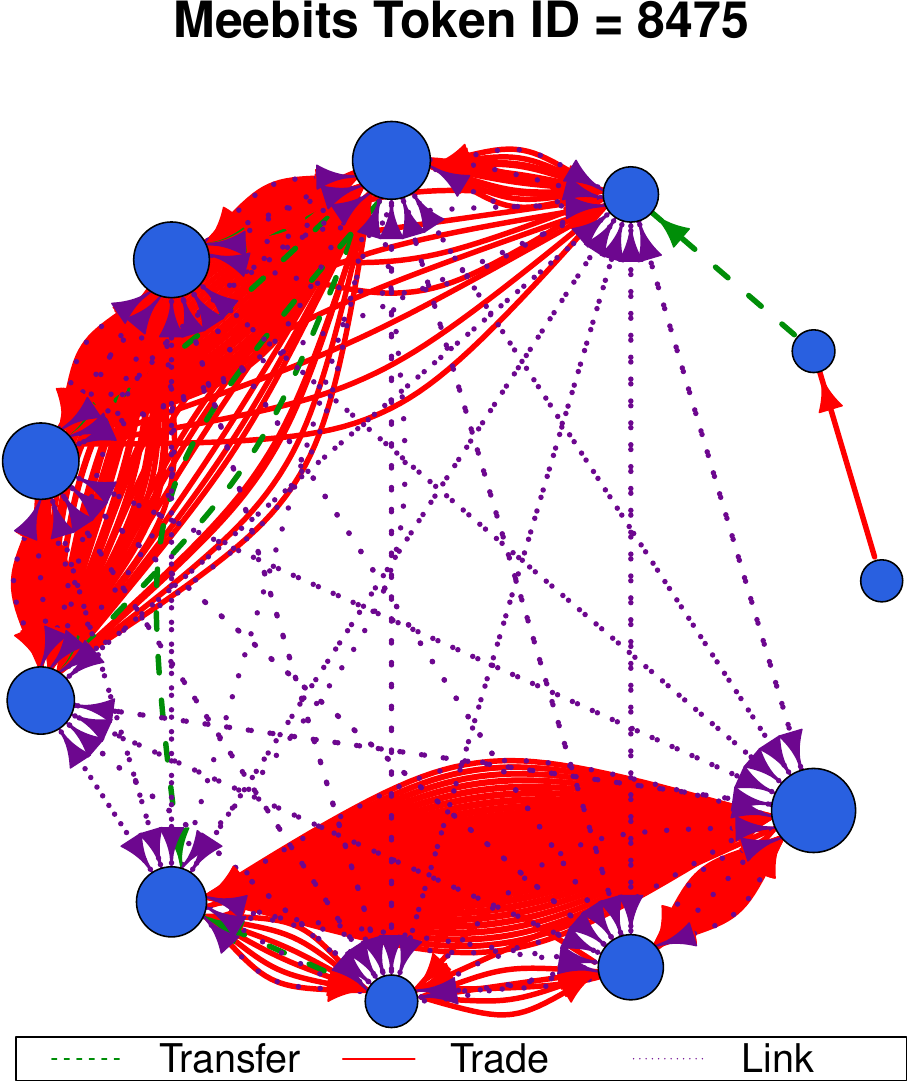}}  &
\subcaptionbox{\label{clonex}}{\includegraphics[width=0.4\textwidth]{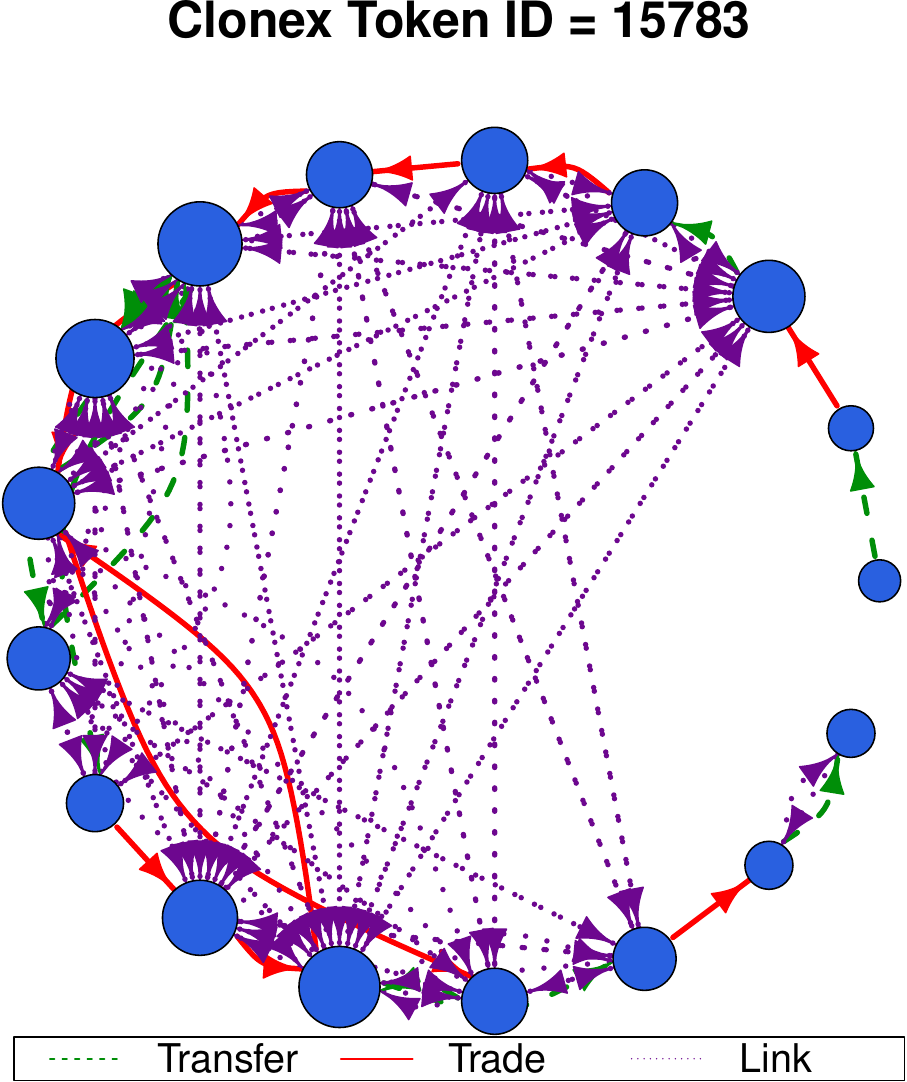}} \\
\subcaptionbox{\label{kennel}}{\includegraphics[width=0.4\textwidth]{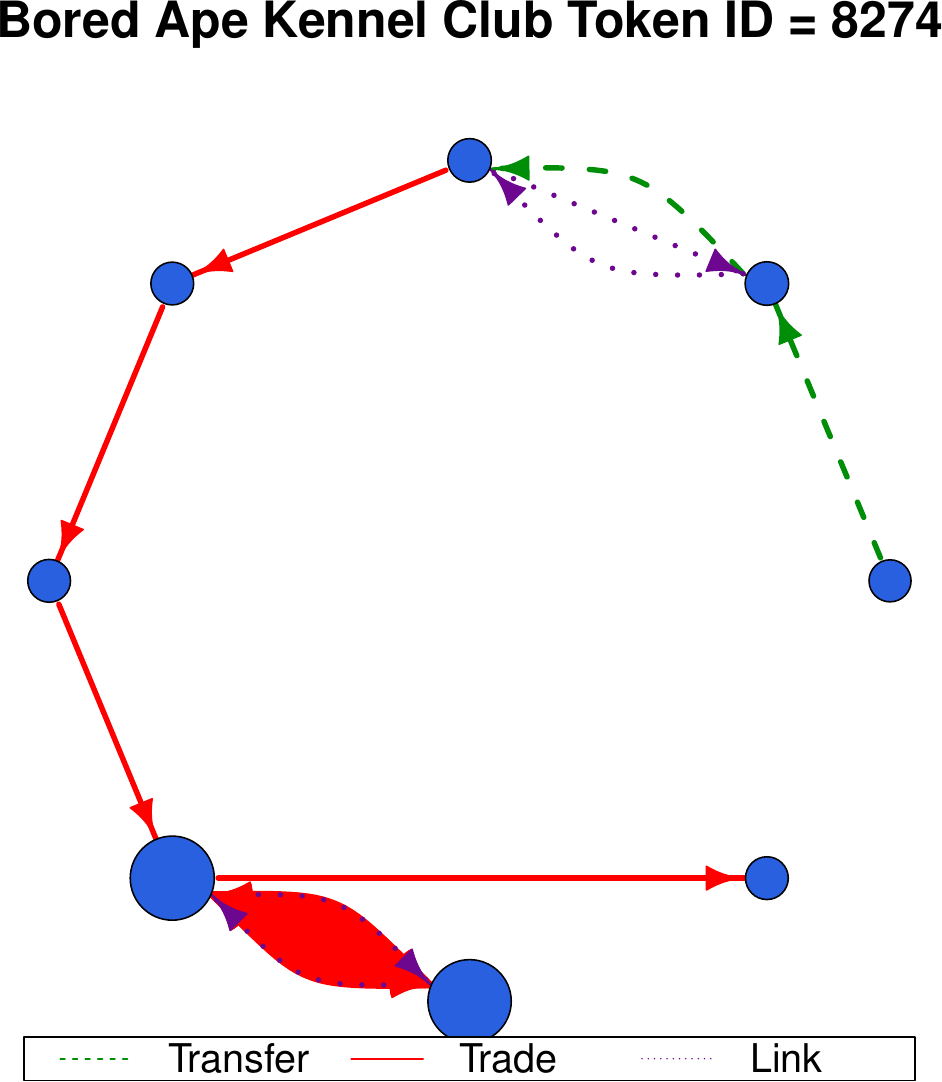}} &
\subcaptionbox{\label{mutant}}{\includegraphics[width=0.4\textwidth]{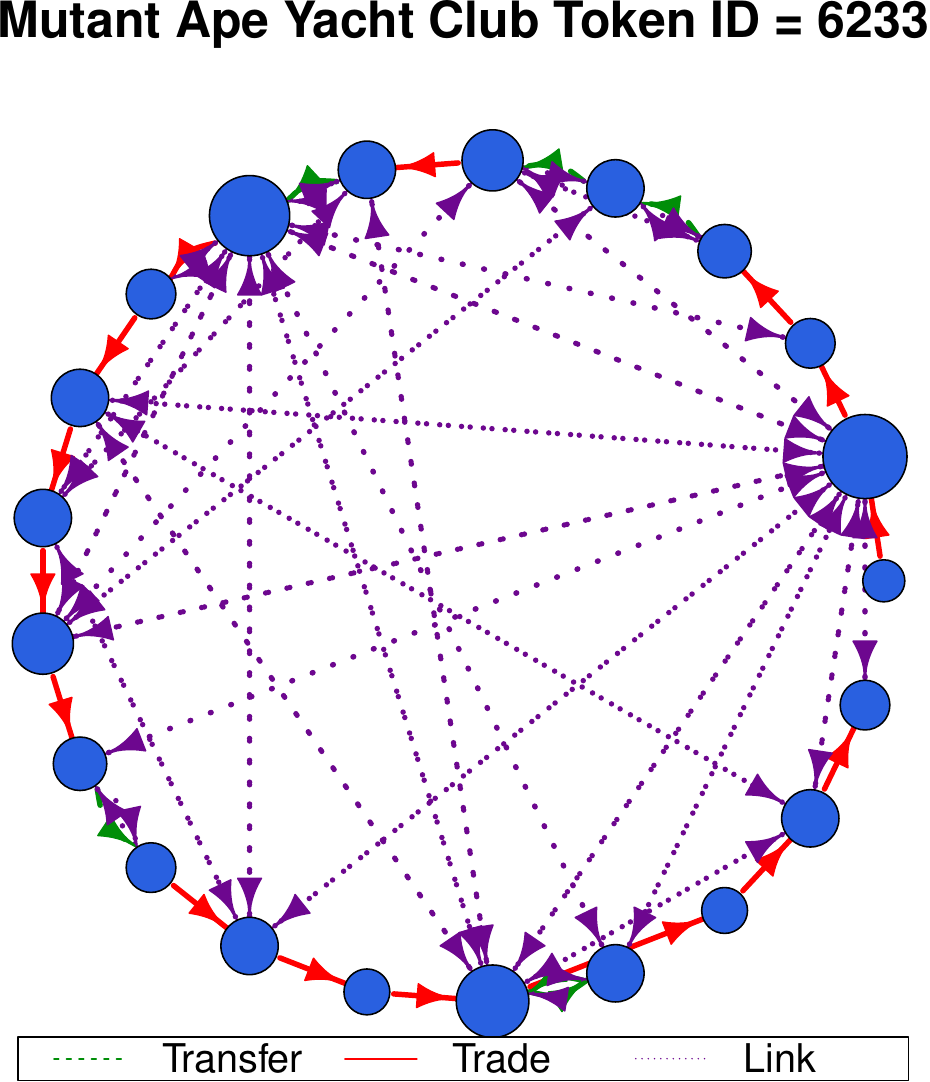}} \\
\end{tabular}
\caption{Handpicked graphs that represent the effectiveness of the proposed method in identifying wash trades. Accounts are represented as vertices while edges are of three types namely, Transfers, Trades or Links.}
\label{graphs}
\end{figure}

In Figure \ref{graphs}, we show six graphs that are highlighted by our algorithm with high scores. The graphs were handpicked among the highest scored within the individual collection to visually show the suspicious behavior. The graphs have three types of edges connecting wallets that have participated in either trading or transferring a particular NFT. Additionally, a dotted line represents edges linked from the Ethereum graph.

The selected graphs display different patterns of wash trading. From Sub figure \ref{azuki} we can observe the typical "Honey Pot" approach where the price of an asset is artificially inflated among well-connected nodes in the Ethereum graphs (purple dotted edges), which creates a false speculative belief that the price increase will continue until a buy is filled by an unconnected wallet. This is evident in the last three nodes executing trades in an attempt to cut their loss. A similar example is shown in Sub figure \ref{clonex} with the exception of one honest account purchase in the beginning, which spiked wash trading.

Perhaps even more evident due to its simplicity is the graph in Sub figure \ref{kennel} where we can observe two connected in the Ethereum graph executed a high number of cyclic trades until finally, an independent account purchased the NFT.

Previous research ~\cite{SERNEELS2023103374} highlighted token $8475$ within the Meebits collection as an example of wash trading. Our algorithm further strengthens this claim by scoring the token in third place within the collection. The high number of cycles in the trading graph among connected accounts can be observed in Sub figure \ref{meebits}. 

Additionally, Table \ref{tab:tokens_stats} shows the selected graphs in Figure \ref{graphs} through a numeric view, which highlights the extent and severity of wash trading in NFT markets.

\begin{table}[]
\begin{tabular}{l|l|l|l|l|l|l}
\textbf{Collection}             & \textbf{Token Id} & \textbf{Total Volume}  & \textbf{Washed Volume} & \textbf{Wash Sales} & \textbf{Total Sales} & \textbf{Ratio} \\ \hline
Kennel Apes & 8274     & \$5,419,892   & \$4,341,469   & 108        & 108   & 0.801 \\ \hline
BAYC  & 7165     & \$877,981     & \$334,449     & 12         & 16    & 0.381 \\ \hline
Mutant Apes & 6233     & \$1,516,586   & \$382,966     & 10         & 16    & 0.253 \\ \hline
Azuki                  & 6233     & \$1,190,944   & \$245,940     & 8          & 13    & 0.207 \\ \hline
Clonex                 & 15783    & \$160,025     & \$90,395      & 8          & 11    & 0.565 \\ \hline
Meebits                & 8475     & \$179,325,909 & \$179,166,551 & 273        & 275   & 0.999 \\ \hline
\end{tabular}
\caption{Output of the wash trade detection algorithm for selected tokens from Figure \ref{graphs}}
\label{tab:tokens_stats}
\end{table}

\begin{figure}[h!]
  \includegraphics[width=0.9\textwidth]{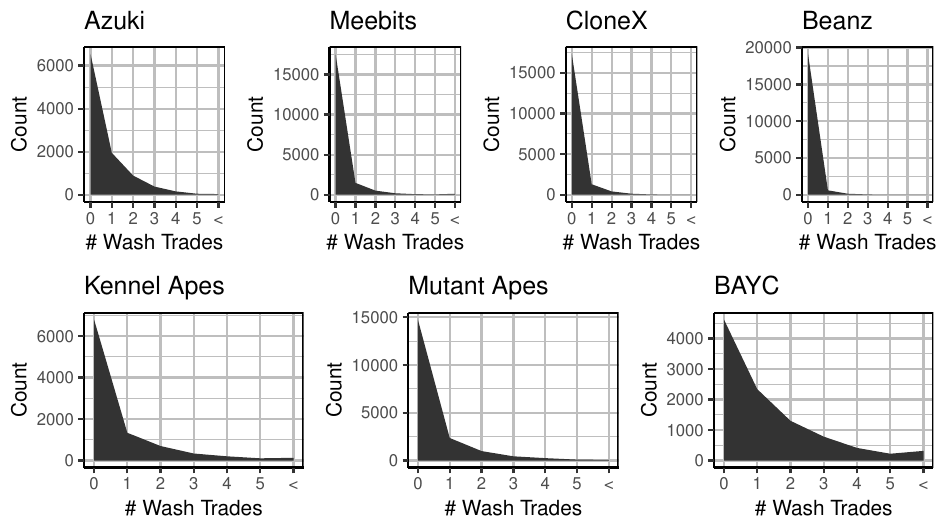}
  \caption{Distribution of wash trades per NFT collection. The symbol < denotes the count of NFTs that were wash traded more than 5 times.}
  \label{fig:distribution}
\end{figure}

From Figure~\ref{fig:distribution}, can be observed that the majority of NFT tokens do not exhibit wash trading activity. Furhtermore, we highlight that BAYC collection stands out by the number of tokens being wash traded, which can be attributed to first mover advantage in popular NFT collections. Moreover, we observe that the wash trading activity is focused on a small selection of tokens within individual collections.

\begin{figure}[h!]
  \includegraphics[width=0.9\textwidth]{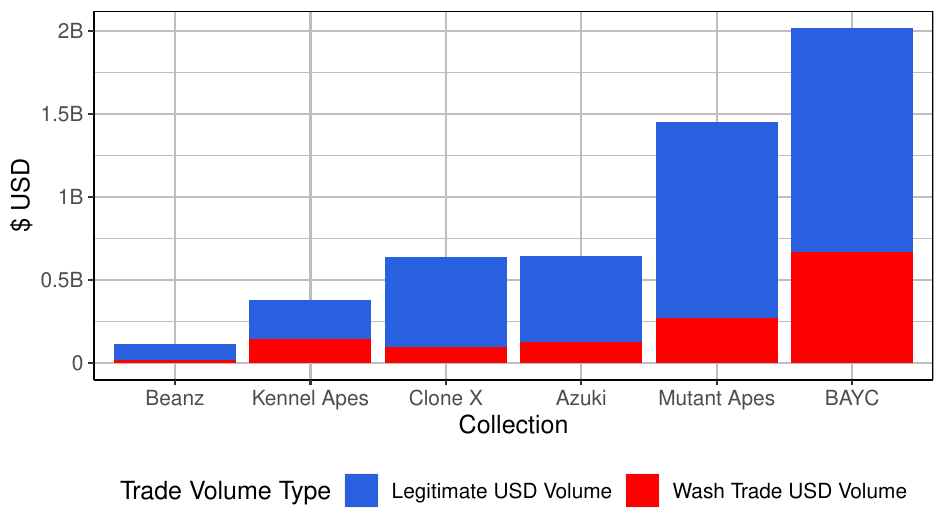}
  \caption{Legitimate total trade volume and total wash trade volume in USD per NFT collection.}
  \label{fig:volume}
\end{figure}

Figure~\ref{fig:volume}, shows a stacked bar chart of the total USD traded volume per collection, distinguishing between legitimate USD volume and the total USD volume flagged as suspicious by our algorithm. 
We have excluded the Meebits collection from the visualization as its overwhelming trade volume distorts the representation.
The Meebits collection stands out for its exceptionally high wash trade activity, with respective total USD volume, legitimate USD volume, and wash trade USD volume at: 9,350B\$, 0,577B\$, 8,77B\$. 
This translates to a staggering 94\% of the Meebits collection's volume attributed to wash trades.
Our observations echo the findings from other research, underscoring the towering wash trading in the Meebits collection~\cite{SERNEELS2023103374}.

Given the overwhelming wash trade activity observed in the Meebits collection, we've opted to exclude it from our estimation of the wash trade volume ratio. This calculation is instead based on the remaining six collections. After this careful consideration, the wash trade ratio for these collections is found to account for a substantial 25.24\% of the total USD volume transacted across the evaluated NFT collections.

\section{Conclusion}
In this paper, we presented an new method for detecting wash trading in decentralized markets. Our approach involves combining the NFT trade graphs with the linkability graphs computed in the underlying Ethereum graph. We have demonstrated that our proposed method is effective in identifying well-established suspicious behavior from relevant literature and is computationally more suitable for analyzing large graphs. Furthermore, with the added information from the linkability network, we are able to detect many more cases previously unfledged. We have showcased the effectiveness of our proposed algorithm through visualizing the trade graphs coupled with the linkability graphs and observing well-known patterns of wash trading effectively scored by our algorithm. Our findings suggest that wash trading in unregulated NFT markets is an underestimated concern and is much more widespread both in terms of frequency as well as volume. The authors believe the contributions presented in this paper will contribute to the development of detection methods as well as help policy makers and regulators protect investors.

\section*{Abbreviations}
\begin{tabular}{ll}
    NFT & Non Fungible Token \\
    DeFi & Decentralized Finance \\
    API & Application Programming Interface \\
    EOA & Externally Owned Account \\
    BFS & Breadth First Search \\
    ETH & Ethereum \\
    DEX & Decentralized Exchange \\
    BAYC & Bored Ape Yacht Club \\
    ETL & Extract Transform Load \\
\end{tabular}
\bibliographystyle{template/bmc-mathphys} 
\bibliography{bmc_article}      


\begin{thebibliography}{12}
\ifx \bisbn   \undefined \def \bisbn  #1{ISBN #1}\fi
\ifx \binits  \undefined \def \binits#1{#1}\fi
\ifx \bauthor  \undefined \def \bauthor#1{#1}\fi
\ifx \batitle  \undefined \def \batitle#1{#1}\fi
\ifx \bjtitle  \undefined \def \bjtitle#1{#1}\fi
\ifx \bvolume  \undefined \def \bvolume#1{\textbf{#1}}\fi
\ifx \byear  \undefined \def \byear#1{#1}\fi
\ifx \bissue  \undefined \def \bissue#1{#1}\fi
\ifx \bfpage  \undefined \def \bfpage#1{#1}\fi
\ifx \blpage  \undefined \def \blpage #1{#1}\fi
\ifx \burl  \undefined \def \burl#1{\textsf{#1}}\fi
\ifx \doiurl  \undefined \def \doiurl#1{\textsf{#1}}\fi
\ifx \betal  \undefined \def \betal{\textit{et al.}}\fi
\ifx \binstitute  \undefined \def \binstitute#1{#1}\fi
\ifx \binstitutionaled  \undefined \def \binstitutionaled#1{#1}\fi
\ifx \bctitle  \undefined \def \bctitle#1{#1}\fi
\ifx \beditor  \undefined \def \beditor#1{#1}\fi
\ifx \bpublisher  \undefined \def \bpublisher#1{#1}\fi
\ifx \bbtitle  \undefined \def \bbtitle#1{#1}\fi
\ifx \bedition  \undefined \def \bedition#1{#1}\fi
\ifx \bseriesno  \undefined \def \bseriesno#1{#1}\fi
\ifx \blocation  \undefined \def \blocation#1{#1}\fi
\ifx \bsertitle  \undefined \def \bsertitle#1{#1}\fi
\ifx \bsnm \undefined \def \bsnm#1{#1}\fi
\ifx \bsuffix \undefined \def \bsuffix#1{#1}\fi
\ifx \bparticle \undefined \def \bparticle#1{#1}\fi
\ifx \barticle \undefined \def \barticle#1{#1}\fi
\ifx \bconfdate \undefined \def \bconfdate #1{#1}\fi
\ifx \botherref \undefined \def \botherref #1{#1}\fi
\ifx \url \undefined \def \url#1{\textsf{#1}}\fi
\ifx \bchapter \undefined \def \bchapter#1{#1}\fi
\ifx \bbook \undefined \def \bbook#1{#1}\fi
\ifx \bcomment \undefined \def \bcomment#1{#1}\fi
\ifx \oauthor \undefined \def \oauthor#1{#1}\fi
\ifx \citeauthoryear \undefined \def \citeauthoryear#1{#1}\fi
\ifx \endbibitem  \undefined \def \endbibitem {}\fi
\ifx \bconflocation  \undefined \def \bconflocation#1{#1}\fi
\ifx \arxivurl  \undefined \def \arxivurl#1{\textsf{#1}}\fi
\csname PreBibitemsHook\endcsname

\bibitem{SERNEELS2023103374}
\begin{barticle}
\bauthor{\bsnm{Serneels}, \binits{S.}}:
\batitle{Detecting wash trading for nonfungible tokens}.
\bjtitle{Finance Research Letters}
\bvolume{52},
\bfpage{103374}
(\byear{2023}).
doi:\doiurl{10.1016/j.frl.2022.103374}
\end{barticle}
\endbibitem

\bibitem{cong2022crypto}
\begin{botherref}
\oauthor{\bsnm{Cong}, \binits{L.W.}},
\oauthor{\bsnm{Li}, \binits{X.}},
\oauthor{\bsnm{Tang}, \binits{K.}},
\oauthor{\bsnm{Yang}, \binits{Y.}}:
Crypto wash trading.
Technical report,
National Bureau of Economic Research
(2022)
\end{botherref}
\endbibitem

\bibitem{dune}
\begin{botherref}
\oauthor{\bsnm{{Dune}}}:
NFT Wash Trading on Ethereum.
\url{https://dune.com/blog/nft-wash-trading-on-ethereum}.
Accessed: 20123-21-08
(2022)
\end{botherref}
\endbibitem

\bibitem{ante2022non}
\begin{barticle}
\bauthor{\bsnm{Ante}, \binits{L.}}:
\batitle{The non-fungible token (nft) market and its relationship with bitcoin
  and ethereum}.
\bjtitle{FinTech}
\bvolume{1}(\bissue{3}),
\bfpage{216}--\blpage{224}
(\byear{2022})
\end{barticle}
\endbibitem

\bibitem{cao2016}
\begin{barticle}
\bauthor{\bsnm{Cao}, \binits{Y.}},
\bauthor{\bsnm{Li}, \binits{Y.}},
\bauthor{\bsnm{Coleman}, \binits{S.}},
\bauthor{\bsnm{Belatreche}, \binits{A.}},
\bauthor{\bsnm{McGinnity}, \binits{T.M.}}:
\batitle{Detecting wash trade in financial market using digraphs and dynamic
  programming}.
\bjtitle{IEEE Transactions on Neural Networks and Learning Systems}
\bvolume{27}(\bissue{11}),
\bfpage{2351}--\blpage{2363}
(\byear{2016}).
doi:\doiurl{10.1109/TNNLS.2015.2480959}
\end{barticle}
\endbibitem

\bibitem{tariq2022suspicious}
\begin{botherref}
\oauthor{\bsnm{Tariq}, \binits{S.A.}},
\oauthor{\bsnm{Sifat}, \binits{I.}}:
Suspicious trading in nonfungible tokens (nfts): Evidence from wash trading.
Available at SSRN 4097642
(2022)
\end{botherref}
\endbibitem

\bibitem{sifat2023suspicious}
\begin{botherref}
\oauthor{\bsnm{Sifat}, \binits{I.}},
\oauthor{\bparticle{van} \bsnm{Donselaar}, \binits{D.}},
\oauthor{\bsnm{Tariq}, \binits{S.A.}}:
Suspicious trading in nonfungible tokens (nfts).
Available at SSRN 4336439
(2023)
\end{botherref}
\endbibitem

\bibitem{vonwachter2022nft}
\begin{botherref}
\oauthor{\bparticle{von} \bsnm{Wachter}, \binits{V.}},
\oauthor{\bsnm{Jensen}, \binits{J.R.}},
\oauthor{\bsnm{Regner}, \binits{F.}},
\oauthor{\bsnm{Ross}, \binits{O.}}:
NFT Wash Trading: Quantifying suspicious behaviour in NFT markets
(2022).
\arxivurl{2202.03866}
\end{botherref}
\endbibitem

\bibitem{liu2023nft}
\begin{botherref}
\oauthor{\bsnm{Liu}, \binits{D.}},
\oauthor{\bsnm{Piccoli}, \binits{F.}},
\oauthor{\bsnm{Chen}, \binits{K.}},
\oauthor{\bsnm{Tang}, \binits{A.}},
\oauthor{\bsnm{Fang}, \binits{V.}}:
Nft wash trading detection.
arXiv preprint arXiv:2305.01543
(2023)
\end{botherref}
\endbibitem

\bibitem{serneels2023}
\begin{barticle}
\bauthor{\bsnm{Serneels}, \binits{S.}}:
\batitle{Detecting wash trading for nonfungible tokens}.
\bjtitle{Finance Research Letters}
\bvolume{52},
\bfpage{103374}
(\byear{2023}).
doi:\doiurl{10.1016/j.frl.2022.103374}
\end{barticle}
\endbibitem

\bibitem{us_wash-trade}
\begin{botherref}
\oauthor{\bsnm{CETC}}:
7 U.S. Code Chapter 1 - COMMODITY EXCHANGES.
\url{https://www.law.cornell.edu/uscode/text/7/chapter-1}.
Accessed: [29/08/2023]
(2023)
\end{botherref}
\endbibitem

\bibitem{pfitzmann2010terminology}
\begin{botherref}
\oauthor{\bsnm{Pfitzmann}, \binits{A.}},
\oauthor{\bsnm{Hansen}, \binits{M.}}:
A terminology for talking about privacy by data minimization: Anonymity,
  unlinkability, undetectability, unobservability, pseudonymity, and identity
  management
(2010)
\end{botherref}
\endbibitem

\end{thebibliography}

\newcommand{\BMCxmlcomment}[1]{}

\BMCxmlcomment{

<refgrp>

<bibl id="B1">
  <title><p>Detecting wash trading for nonfungible tokens</p></title>
  <aug>
    <au><snm>Serneels</snm><fnm>S</fnm></au>
  </aug>
  <source>Finance Research Letters</source>
  <pubdate>2023</pubdate>
  <volume>52</volume>
  <fpage>103374</fpage>
  <url>https://www.sciencedirect.com/science/article/pii/S1544612322005517</url>
</bibl>

<bibl id="B2">
  <title><p>Crypto wash trading</p></title>
  <aug>
    <au><snm>Cong</snm><fnm>LW</fnm></au>
    <au><snm>Li</snm><fnm>X</fnm></au>
    <au><snm>Tang</snm><fnm>K</fnm></au>
    <au><snm>Yang</snm><fnm>Y</fnm></au>
  </aug>
  <pubdate>2022</pubdate>
</bibl>

<bibl id="B3">
  <title><p>NFT Wash Trading on Ethereum</p></title>
  <aug>
    <au><cnm>{Dune}</cnm></au>
  </aug>
  <source>\url{https://dune.com/blog/nft-wash-trading-on-ethereum}</source>
  <pubdate>2022</pubdate>
  <note>Accessed: 20123-21-08</note>
</bibl>

<bibl id="B4">
  <title><p>The non-fungible token (NFT) market and its relationship with
  Bitcoin and Ethereum</p></title>
  <aug>
    <au><snm>Ante</snm><fnm>L</fnm></au>
  </aug>
  <source>FinTech</source>
  <publisher>MDPI</publisher>
  <pubdate>2022</pubdate>
  <volume>1</volume>
  <issue>3</issue>
  <fpage>216</fpage>
  <lpage>-224</lpage>
</bibl>

<bibl id="B5">
  <title><p>Detecting Wash Trade in Financial Market Using Digraphs and Dynamic
  Programming</p></title>
  <aug>
    <au><snm>Cao</snm><fnm>Y</fnm></au>
    <au><snm>Li</snm><fnm>Y</fnm></au>
    <au><snm>Coleman</snm><fnm>S</fnm></au>
    <au><snm>Belatreche</snm><fnm>A</fnm></au>
    <au><snm>McGinnity</snm><fnm>TM</fnm></au>
  </aug>
  <source>IEEE Transactions on Neural Networks and Learning Systems</source>
  <pubdate>2016</pubdate>
  <volume>27</volume>
  <issue>11</issue>
  <fpage>2351</fpage>
  <lpage>2363</lpage>
</bibl>

<bibl id="B6">
  <title><p>Suspicious trading in nonfungible tokens (nfts): Evidence from wash
  trading</p></title>
  <aug>
    <au><snm>Tariq</snm><fnm>SA</fnm></au>
    <au><snm>Sifat</snm><fnm>I</fnm></au>
  </aug>
  <source>Available at SSRN 4097642</source>
  <pubdate>2022</pubdate>
</bibl>

<bibl id="B7">
  <title><p>Suspicious Trading in Nonfungible Tokens (NFTs)</p></title>
  <aug>
    <au><snm>Sifat</snm><fnm>I</fnm></au>
    <au><snm>Donselaar</snm><fnm>D</fnm></au>
    <au><snm>Tariq</snm><fnm>SA</fnm></au>
  </aug>
  <source>Available at SSRN 4336439</source>
  <pubdate>2023</pubdate>
</bibl>

<bibl id="B8">
  <title><p>NFT Wash Trading: Quantifying suspicious behaviour in NFT
  markets</p></title>
  <aug>
    <au><snm>Wachter</snm><fnm>V</fnm></au>
    <au><snm>Jensen</snm><fnm>JR</fnm></au>
    <au><snm>Regner</snm><fnm>F</fnm></au>
    <au><snm>Ross</snm><fnm>O</fnm></au>
  </aug>
  <pubdate>2022</pubdate>
</bibl>

<bibl id="B9">
  <title><p>NFT Wash Trading Detection</p></title>
  <aug>
    <au><snm>Liu</snm><fnm>D</fnm></au>
    <au><snm>Piccoli</snm><fnm>F</fnm></au>
    <au><snm>Chen</snm><fnm>K</fnm></au>
    <au><snm>Tang</snm><fnm>A</fnm></au>
    <au><snm>Fang</snm><fnm>V</fnm></au>
  </aug>
  <source>arXiv preprint arXiv:2305.01543</source>
  <pubdate>2023</pubdate>
</bibl>

<bibl id="B10">
  <title><p>Detecting wash trading for nonfungible tokens</p></title>
  <aug>
    <au><snm>Serneels</snm><fnm>S</fnm></au>
  </aug>
  <source>Finance Research Letters</source>
  <pubdate>2023</pubdate>
  <volume>52</volume>
  <fpage>103374</fpage>
  <url>https://www.sciencedirect.com/science/article/pii/S1544612322005517</url>
</bibl>

<bibl id="B11">
  <title><p>7 U.S. Code Chapter 1 - COMMODITY EXCHANGES</p></title>
  <aug>
    <au><cnm>CETC</cnm></au>
  </aug>
  <source>\url{https://www.law.cornell.edu/uscode/text/7/chapter-1}</source>
  <pubdate>2023</pubdate>
  <note>Accessed: [29/08/2023]</note>
</bibl>

<bibl id="B12">
  <title><p>A terminology for talking about privacy by data minimization:
  Anonymity, unlinkability, undetectability, unobservability, pseudonymity, and
  identity management</p></title>
  <aug>
    <au><snm>Pfitzmann</snm><fnm>A</fnm></au>
    <au><snm>Hansen</snm><fnm>M</fnm></au>
  </aug>
  <pubdate>2010</pubdate>
</bibl>

</refgrp>
} 


\end{document}